\documentclass[a4paper,11pt]{article}
\pdfoutput=1
\usepackage{jheppub} 
\usepackage{mathtools}
\newcommand{\Hrh}{H_\text{rh}}
\newcommand{\Trh}{T_\text{rh}}
\newcommand{\gs}{g_\star}
\newcommand{\gss}{g_{\star s}}
\newcommand{\arh}{a_\text{rh}}
\newcommand{\rp}{\rho_\phi}
\newcommand{\rR}{\rho_R}
\newcommand{\Br}{\text{Br}}
\newcommand{\mrh}{{m_\phi^\text{rh}}}

\title{Dark Matter Ultraviolet Freeze-in\\in General Reheating Scenarios}
\author{Nicolás Bernal,}
\author{Kuldeep Deka}
\author{and Marta Losada}
\affiliation{New York University Abu Dhabi\\
PO Box 129188, Saadiyat Island, Abu Dhabi, United Arab Emirates}

\emailAdd{nicolas.bernal@nyu.edu}
\emailAdd{kuldeep.deka@nyu.edu}
\emailAdd{marta.losada@nyu.edu}

\abstract{The dynamics of cosmic reheating, that is, on how the energy stored in the inflaton is transferred to the standard model (SM) thermal bath, is largely unknown. In this work, we show that the phenomenology of the nonbaryonic dark matter (DM) ultraviolet freeze-in production strongly depends on the dynamics of the cosmic-reheating era. Using a general parametrization for the Hubble expansion rate and SM temperature, we thoroughly investigate DM production during reheating, not only recovering earlier findings that focused on specific cases, but also exploring alternative scenarios. Additionally, we derive a generalized framework for DM production via inflaton decays and identify the viable parameter space, while simultaneously addressing constraints from CMB observations. As illustrative examples, we explore gravitational DM production through scatterings of SM particles or inflatons, deriving well-defined parameter regions for these scenarios.}

\begin{document} 
\begin{flushright}
\end{flushright}
\maketitle
\flushbottom

\section{Introduction}
Substantial astrophysical and cosmological data robustly support the existence of nonbaryonic dark matter (DM) in the Universe. To be viable, a DM candidate must fulfill several requirements: it needs to be electromagnetically neutral, cosmologically stable, and non-relativistic at the onset of the Big Bang nucleosynthesis era. It should also account for a relic density of $\Omega h^2 \simeq 0.12$, corresponding to 27\% of the total energy content of the Universe~\cite{Planck:2018vyg}; for a comprehensive review, see Ref.~\cite{Cirelli:2024ssz}.

In the early Universe, the weakly interacting massive particle (WIMP) paradigm has been the most popular DM production mechanism. In this scenario, DM has mass on the electroweak scale and couples to the standard model (SM) thermal plasma with rates typical of electroweak interactions. WIMPs attain thermal equilibrium with the SM plasma and later experience chemical freeze-out, giving rise to the observed DM relic abundance. The WIMP approach is particularly appealing because it can be probed with several techniques, including direct detection, indirect detection, and collider experiments. However, the lack of experimental evidence and the stringent constraints imposed on the expected parameter space prompt the consideration of alternatives beyond the standard WIMP paradigm~\cite{Arcadi:2017kky, Roszkowski:2017nbc, Arcadi:2024ukq}.

In contrast to WIMPs, feebly interacting massive particles (FIMPs) couple to the visible sector very feebly, therefore evading the current experimental constraints~\cite{McDonald:2001vt, Choi:2005vq, Kusenko:2006rh, Petraki:2007gq, Hall:2009bx, Bernal:2017kxu}. The FIMP framework demands significantly reduced interaction rates between the dark and visible sectors. This can be accomplished in its infrared variant through very small couplings (usually around $\sim 10^{-11}$), or in its ultraviolet (UV) form via non-renormalizable operators, suppressed by a large energy scale~\cite{Elahi:2014fsa}. The latter scenario is particularly interesting, as the DM yield is sensitive to the highest temperature reached by the SM plasma~\cite{Giudice:2000ex}, and, in particular, to the cosmic reheating era~\cite{Kofman:1997yn}.

In general, there are huge uncertainties on the history of the Universe before the Big Bang nucleosynthesis epoch~\cite{Allahverdi:2020bys, Batell:2024dsi}. In particular, the dynamics of cosmic reheating, that is, on how the energy stored in the inflaton is transferred to the SM thermal bath, is largely unknown~\cite{Hasegawa:2019jsa}. During the reheating phase, the Hubble expansion of the Universe was dominated by the inflaton. Its energy density is generally considered to evolve like non-relativistic matter or radiation, which are associated with an inflaton oscillating at the bottom of a quadratic or quartic potential, respectively. However, it can also scale faster than radiation, as in the case of kination~\cite{Spokoiny:1993kt, Ferreira:1997hj}, or even more faster, as in the context of ekpyrotic~\cite{Khoury:2001wf, Khoury:2003rt} or cyclic models~\cite{Gasperini:2002bn, Erickson:2003zm, Barrow:2010rx, Ijjas:2019pyf}, with further discussion in Ref.~\cite{Scherrer:2022nnz}. Moreover, the inflaton could decay or annihilate into various types of SM particles, leading to different dependencies of the temperature of the SM bath on the cosmic scale factor. In particular, during reheating, the SM temperature could increase, decrease, or even stay constant~\cite{Giudice:2000ex, Co:2020xaf}.

In this work, we demonstrate that the phenomenology of the DM UV freeze-in production strongly depends on the dynamics of the cosmic reheating era. Using a general parametrization for the Hubble expansion rate and SM temperature, we carefully study the DM production during reheating. This allows us not only to recover previous results reported in the literature, but also to systematically explore large unexplored regions of the parameter space of reheating. In particular, we study the DM UV freeze-in production from decays of the inflaton, and annihilations parametrized by general temperature dependence. Notably, the most general parameterization for inflaton-decay scenarios has not been addressed in prior studies. As motivated examples, we specifically studied the cases of gravitational DM production from the scattering of SM particles or inflatons~\cite{Ema:2015dka, Garny:2015sjg, Tang:2016vch, Ema:2016hlw, Tang:2017hvq, Garny:2017kha, Bernal:2018qlk, Ema:2018ucl, Almeida:2018oid, Ema:2019yrd, Chianese:2020yjo, Bernal:2020bfj, Ahmed:2020fhc, Bernal:2020qyu, Chianese:2020khl, Kolb:2020fwh, Bernal:2020ili, Redi:2020ffc, Ling:2021zlj, Mambrini:2021zpp, Bernal:2021kaj, Barman:2021ugy, Barman:2021qds, Haque:2021mab, Clery:2021bwz, Barman:2022tzk, Clery:2022wib, Garcia:2022vwm, Kaneta:2022gug, Basso:2022tpd, Barman:2022qgt, Haque:2023yra, Kolb:2023dzp, Garcia:2023awt, Garcia:2023qab, Kaneta:2023uwi, Barman:2023opy, Garcia:2023obw, Kolb:2023ydq, Racco:2024aac, Dorsch:2024nan, Choi:2024bdn, Verner:2024agh, Jenks:2024fiu, Lee:2024wes, Henrich:2024rux, Xu:2024cey, Maleknejad:2024hoz}.

The paper is organized as follows. In Section~\ref{sec:reheating} the parametrization of the cosmic reheating era is introduced. In Section~\ref{sec:dm} the DM UV freeze-in production is studied, after and during reheating, using a general parametrization for the collision term. In addition, DM genesis through direct decays of the inflaton is also investigated. As an example, in Section~\ref{sec:particles} the cases of gravitational DM production from the scattering of SM particles or inflatons are analyzed. Finally, we summarize our findings in Section~\ref{sec:conclusion}.

\section{Parametrizing Cosmic Reheating} \label{sec:reheating}
The exact nature of the reheating phase remains uncertain, but is commonly modeled as being driven by the decay or annihilation of a scalar field $\phi$, often identified with the inflaton. This process is characterized by an effective equation-of-state parameter $\omega$, such that the energy density of the scalar field evolves as $\rho_\phi(a) \propto a^{-3(1+\omega)}$ during reheating. Reheating transfers energy from $\phi$ to the Standard Model (SM) radiation, with the radiation energy density expressed as
\begin{equation}
    \rR(T) \equiv \frac{\pi^2}{30}\, \gs(T)\, T^4,
\end{equation}
where where $\gs(T)$ is the effective number of relativistic degrees of freedom contributing to the radiation energy~\cite{Drees:2015exa}, and $T$ is the temperature of the SM thermal bath. The reheating temperature $\Trh$ is defined as the temperature at which the radiation energy density is equal to the inflaton energy density, $\rp (\arh) = \rR (\arh) = 3\, \Hrh^2\, M_P^2$, where $\arh$ is the scale factor at $T = \Trh$, $\Hrh \equiv H(\arh)$ is the Hubble expansion rate at that time, and $M_P \simeq 2.4 \times 10^{18}$~GeV is the reduced Planck mass. It follows that the Hubble rate is~\cite{Bernal:2024yhu, Bernal:2024jim, Bernal:2024ndy}
\begin{equation} \label{eq:Hubble}
    H(a) = \sqrt{\frac{\rp + \rR}{3\, M_P^2}} = \Hrh \times
    \begin{dcases}
        \left(\frac{\arh}{a}\right)^\frac{3(1+\omega)}{2} &\text{ for } a \leq \arh,\\
        \left(\frac{\gs(T)}{\gs(\Trh)}\right)^\frac12 \left(\frac{\gss(\Trh)}{\gss(T)}\right)^\frac23 \left(\frac{\arh}{a}\right)^2 &\text{ for } \arh \leq a\,,
    \end{dcases}
\end{equation}
where $\gss(T)$ denotes the effective number of relativistic degrees of freedom contributing to the entropy density $s$~\cite{Drees:2015exa} defined as
\begin{equation}
    s(T) = \frac{2 \pi^2}{45}\, \gss(T)\, T^3.
\end{equation}

During reheating, the inflaton decays or annihilates, transferring its energy density to the SM thermal bath. The evolution of $\rp$ and the SM energy density $\rR$ can be tracked with the Boltzmann equations\footnote{If $\phi$ is coherently oscillating in a monomial potential $V(\phi) \propto \phi^p$, the right hand side of Eqs.~\eqref{eq:BErp} and~\eqref{eq:BErR} should contain an extra multiplicative factor $2\, p/(2+p) = 1+\omega$~\cite{Turner:1983he, Allahverdi:2020bys}.}
\begin{align}
    \frac{d\rp}{dt} + 3 (1+\omega)\, H\, \rp &= - \Gamma\, \rp\,, \label{eq:BErp}\\
    \frac{d\rR}{dt} + 4\, H\, \rR &= + \Gamma\, \rp\,, \label{eq:BErR}
\end{align}
where $\Gamma(a)$ corresponds to an effective decay width of the inflaton, which by using  Eq.~\eqref{eq:BErR} can be parametrized as
\begin{equation} \label{eq:Gamma}
    \Gamma(a) = 4\, (1 - \alpha)\, \Hrh \left(\frac{\arh}{a}\right)^\frac{8\alpha - 3(\omega+1)}{2},
\end{equation}
for $\alpha \leq 1$. The dependence on the scale factor in Eq.~\eqref{eq:Gamma} has been chosen so that~\cite{Bernal:2024yhu, Bernal:2024jim, Bernal:2024ndy}
\begin{equation}\label{eq:Tem}
    T(a) = \Trh \times
    \begin{dcases}
        \left(\frac{\arh}{a}\right)^\alpha &\text{ for } a_I \leq a \leq \arh,\\
        \left(\frac{\gss(\Trh)}{\gss(T)}\right)^\frac13 \frac{\arh}{a} &\text{ for } \arh \leq a\,,
    \end{dcases}
\end{equation}
where $T_I \equiv T(a_I) =  \Trh \times (\arh/a_I)^\alpha$. Interestingly, during the reheating period the Universe can reach a temperature higher than $\Trh$ if $\alpha > 0$~\cite{Giudice:2000ex}. However, the temperature can remain constant if $\alpha = 0$, or even increase in scenarios where $\alpha < 0$~\cite{Co:2020xaf}.

After reheating, the Universe transitions to a radiation-dominated phase, where $H(a) \propto a^{-2}$ and $T(a) \propto a^{-1}$. Furthermore, we note that, as expected, Eq.~\eqref{eq:Gamma} reproduces the standard case of a constant decay width in the cases $\omega = 0$ and $\alpha = 3/8$, and implies a vanishing decay width if $\alpha = 1$ as in the case of kination~\cite{Spokoiny:1993kt, Ferreira:1997hj}.

The BICEP/Keck bound on the tensor-to-scalar ratio implies that the Hubble parameter during inflation $H_I \equiv H(a_I)$ is bounded from above $H_I \leq 2.0 \times 10^{-5}~M_P$~\cite{BICEP:2021xfz}. That translates into a bound on the temperature $T_I$ reached by the thermal bath at the beginning of reheating given by~\cite{Bernal:2024yhu}
\begin{equation} \label{eq:TI}
    T_I \simeq \Trh \left[\frac{90}{\pi^2\, \gs(\Trh)}\, \frac{H_I^2\, M_P^2}{\Trh^4}\right]^\frac{\alpha}{3 (1+\omega)}.
\end{equation}

We provide some remarks on the parametrization of reheating. The most commonly studied scenario involves a heavy inflaton whose energy density evolves like non-relativistic matter ($\omega = 0$) and undergoes perturbative decay into pairs of Standard Model (SM) particles. This leads to a temperature scaling given by $\alpha = 3/8$~\cite{Giudice:2000ex}. This scenario can be realized in viable inflationary models like Starobinsky inflation\cite{Starobinsky:1980te} or polynomial inflation~\cite{Drees:2021wgd, Bernal:2021qrl, Drees:2022aea, Bernal:2024ykj}.  However, alternative scenarios are also possible. For example, in the framework of $\alpha$-attractor inflation models~\cite{Kallosh:2013hoa, Kallosh:2013maa}, the inflaton can oscillate near the minimum of a monomial potential $V(\phi) \propto \phi^p$ with $p \geq 2$ during reheating, with the corresponding equation-of-state parameter given by $\omega = (p-2)/(p+2)$~\cite{Turner:1983he}. The value of $\alpha$ depends on the mechanism of energy transfer from the inflaton to the radiation, which is determined by the nature of the inflaton-matter interactions~\cite{Co:2020xaf, Garcia:2020wiy, Xu:2023lxw, Barman:2024mqo}. For example, if the inflaton decays into bosons, $\alpha = 3/(2(p+2))$, while for decays into fermions with $p < 7$, $\alpha = 3(p-1)/(2(p+2))$. In the case of fermionic inflaton decay and $p > 7$, one has $\alpha = 1$.  In scenarios where the inflaton annihilates into bosons via contact interactions, $\alpha = 9/(2(p+2))$ for $p \geq 3$. For reheating through $s$-channel annihilation mediated by a light scalar, resonant effects yield $\alpha = 3(7-2p)/(2(p+2))$ or $\alpha = 3(5-p)/(2(p+2))$ for bosonic and fermionic final states, respectively~\cite{Barman:2024mqo}. If the mediator is massive, the inflaton annihilation into fermion pairs results in $\alpha = 1$~\cite{Barman:2024mqo}. In addition, there are scenarios where the temperature remains constant during reheating, corresponding to $\alpha = 0$~\cite{Co:2020xaf, Barman:2022tzk, Chowdhury:2023jft, Cosme:2024ndc}. If the inflaton energy density decreases faster than free radiation (i.e., $\omega > 1/3$), the inflaton may not need to decay or annihilate completely, and one can have $\alpha = 1$, as in kination~\cite{Spokoiny:1993kt, Ferreira:1997hj}.

\begin{figure}[t!]
    \def\sepf{0.496}
    \centering
    \includegraphics[width=\sepf\columnwidth]{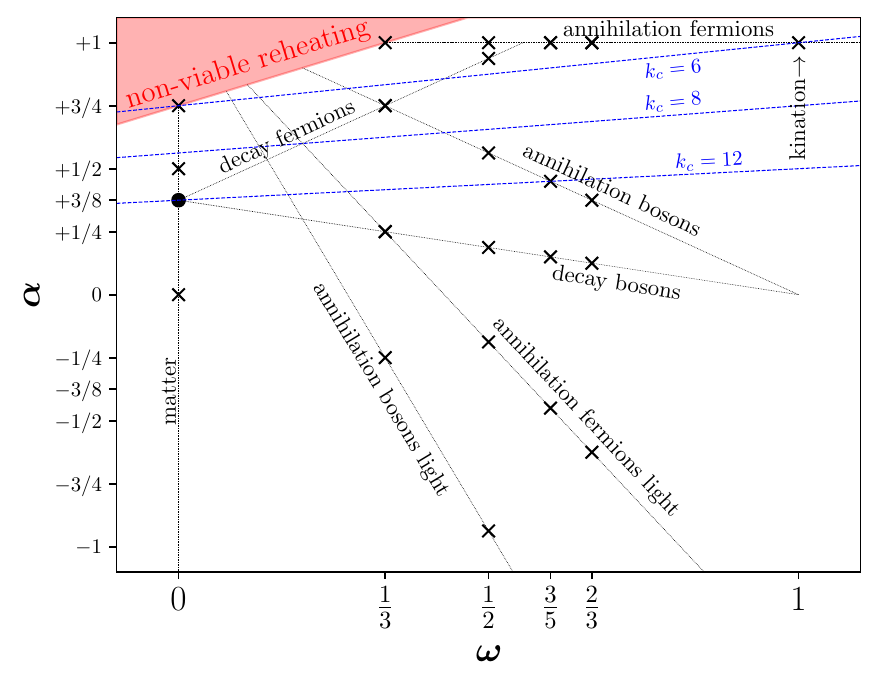}
    \caption{Summary of the different reheating scenarios. The black dot corresponds to the standard case where the inflaton scales as non-relativistic matter and decays into SM particles with a constant decay width, while the black crosses correspond to the alternative scenarios described in the text. The red area in the upper left corner does not give rise to viable reheating. The central blue dashed line corresponds to $k_c \equiv \frac32 \frac{\omega + 3}{\alpha} = 8$; above, over and below it, the gravitational DM production from SM scatterings has an power-law, logarithmic or order-one enhancement during reheating, respectively. For reference, the lines $k_c = 6$ and $k_c = 12$ are also shown.}
    \label{fig:alpha_omega}
\end{figure}

These diverse reheating scenarios are illustrated in Fig.\ref{fig:alpha_omega} in the [$\omega$, $\alpha$] plane. The black lines represent specific scenarios, highlighting how $\omega$ and $\alpha$ are related to the reheating dynamics. The vertical gray dotted line corresponds to the $\omega = 0$ case. The red region in the upper left, where $\alpha > 3(1+\omega)/4$, represents unphysical configurations in which the SM radiation energy density never exceeds the inflaton energy density~\cite{Co:2020xaf}.

\section{UV Dark Matter Production} \label{sec:dm}
The evolution of the DM number density $n$ in the early Universe can be tracked with the Boltzmann equation
\begin{equation} \label{eq:BE0}
    \frac{dn}{dt} + 3\, H\, n = \gamma\,,
\end{equation}
where $\gamma$ corresponds to the DM production rate density, typically containing a contribution $\gamma_a$ from annihilation, and a contribution $\gamma_d$ from decays:
\begin{equation}
    \gamma = \gamma_a + \gamma_d\,.
\end{equation}
In the general case where the SM entropy is not conserved, it is convenient to rewrite Eq.~\eqref{eq:BE0} as a function of the comoving number density $N \equiv n\, a^3$ as
\begin{equation} \label{eq:BE}
    \frac{dN}{da} = \frac{a^2}{H}\, \gamma\,.
\end{equation}

The UV DM production rate density from annihilations is assumed to have a strong temperature dependence, and is parametrized as
\begin{equation} \label{eq:gammaa}
    \gamma_a(T) = \frac{T^k}{\Lambda^{k-4}}\,,
\end{equation}
where $k > 5$ being a constant and $\Lambda$ corresponds to an effective energy scale of new physics. For this approach to be valid, $\Lambda$ must be much higher than the temperatures reached by the SM thermal bath and the DM mass. Concrete examples of the rate $\gamma_a$ will be presented in Section~\ref{sec:particles}.

Furthermore, DM can also be produced directly from decays. For example, during reheating, the inflaton with mass $m_\phi$ can also directly decay into $x$ DM particles, with a small branching fraction $\Br$, if $m_\phi > x\, m$.\footnote{Even if the inflaton does not directly couple to DM particles, loop-induced decays through operators like the one in Eq.~\eqref{eq:gammaa} cannot be avoided~\cite{Kaneta:2019zgw}.} The corresponding interaction rate density for DM production is
\begin{equation}
    \gamma_d = \mathcal{C}\, \Gamma\, \rp\,,
\end{equation}
with $\mathcal{C} \equiv \frac{x\, \Br}{m_\phi}$, and $\Gamma(a)$ given in Eq.~\eqref{eq:Gamma}. Therefore,
\begin{equation} \label{eq:dec}
    \gamma_d = 12\, (1-\alpha)\, \mathcal{C}\, M_P^2\, \Hrh^3 \left(\frac{\arh}{a}\right)^\frac{8\alpha + 3(1 + \omega)}{2}.
\end{equation}
It is interesting to note that, given the parametric form of Eq.~\eqref{eq:dec}, the right-hand side of Eq.~\eqref{eq:BE} becomes independent of the equation-of-state parameter $\omega$.

In nonquadratic potentials, the inflaton mass is expected to be a field-dependent quantity that can be written as a function of the scale factor as
\begin{equation} \label{eq:mrh}
    m_\phi(a) = \mrh \left(\frac{\arh}{a}\right)^{3\, \omega},
\end{equation}
where $\mrh$ corresponds to the inflaton mass at the end of reheating, when $a = \arh$; see e.g. Ref.~\cite{Garcia:2020wiy, Barman:2023rpg, Barman:2024mqo}. In this case, it is convenient to generalize $\mathcal{C} \equiv \frac{x\, \Br}{\mrh}$. For a field-dependent inflaton mass, the decay rate density becomes
\begin{equation}
    \gamma_d = 12\, (1-\alpha)\, \mathcal{C}\, M_P^2\, \Hrh^3 \left(\frac{\arh}{a}\right)^\frac{8\alpha + 3(1 - \omega)}{2}.
\end{equation}
Once the inflaton mass is not constant (that is, for $\omega \ne 0$), it has to be guaranteed that, at all times, $m_\phi(a) < M_P$, which translates into
\begin{align}
    \frac{T_I}{\Trh} < \left(\frac{M_P}{\mrh}\right)^\frac{\alpha}{3 \omega} \quad \text{ for } \quad \alpha > 0\,, \label{eq:TImax}\\
    \frac{T_I}{\Trh} > \left(\frac{M_P}{\mrh}\right)^\frac{\alpha}{3 \omega} \quad \text{ for } \quad \alpha < 0\,, \label{eq:TImin}
\end{align}
which implies an upper (lower) bound for $T_I$ if $\alpha > 0$ ($\alpha < 0$).

In the following subsections, the production of DM during $i)$ the SM-radiation era (that is, after the end of reheating) and $ii)$ the reheating era will be studied separately.

\subsection{After Reheating} \label{sec:after}
In the SM radiation-dominated era, after the end of reheating, the DM abundance produced by annihilation of SM particles with a rate give by Eq.~\eqref{eq:gammaa} can be computed by integrating Eq.~\eqref{eq:BE} between $\arh$ and $a_0$, so that
\begin{equation}
    N_0 = \int_{\arh}^{a_0} \frac{a^2}{H}\, \gamma_a\, da \simeq \frac{1}{k-5}\, \frac{\Trh^k}{\Hrh\, \Lambda^{k-4}}\, \arh^3\,,
\end{equation}
for $k > 5$, and ignoring for now any DM abundance produced before or during reheating. It follow that the DM number density at late times (when $T = T_0$) is
\begin{equation}
    n_0 \simeq \frac{3}{(k-5)\, \pi} \sqrt{\frac{10}{\gs(\Trh)}}\, \frac{M_P\, \Trh^{k-2}}{\Lambda^{k-4}} \left(\frac{T_0}{\Trh}\right)^3,
\end{equation}
and therefore the DM yield defined as $Y \equiv n/s$ is
\begin{equation} \label{eq:DMrd}
    Y_0 \simeq \frac{135}{2 (k-5)\, \pi^3\, \gss(T_0)} \sqrt{\frac{10}{\gs(\Trh)}}\, \frac{M_P\, \Trh^{k-5}}{\Lambda^{k-4}}\,.
\end{equation}
Finally, we emphasize that, in the present scenario, possible thermalization and number-changing processes within the dark sector are ignored. They can have a strong impact, in particular by enhancing or even reducing the DM relic abundance by several orders of magnitude~\cite{Bernal:2020gzm, Bhattiprolu:2023akk, Bhattiprolu:2024dmh}. For the sake of simplicity, we assume here that there are no sizable self-interactions within the dark sector.

To match the entire observed DM relic density, it is required that
\begin{equation}
    m\, Y_0 = \frac{\Omega h^2\, \rho_c}{s_0\, h^2} \simeq 4.3 \times 10^{-10}~\text{GeV},
\end{equation}
where $m$ is the DM mass, $Y_0$ is the asymptotic value of the DM yield at low temperatures, $s_0 \simeq 2.69 \times 10^3$~cm$^{-3}$ is the present entropy density~\cite{ParticleDataGroup:2024cfk}, $\rho_c \simeq 1.05 \times 10^{-5}~h^2$~GeV/cm$^3$ is the critical energy density of the Universe, and $\Omega h^2 \simeq 0.12$ is the observed DM relic abundance~\cite{Planck:2018vyg}.

Before closing this section, we emphasize that the previous discussion only applies to the production of DM particles lighter than the reheating temperature $m < \Trh$. Heavier DM particles with mass $m > \Trh$ could also be produced after reheating, but their production rate density in Eq.~\eqref{eq:gammaa} must be multiplied by a Boltzmann suppression factor $\exp[-m/T]$, reflecting the fact that only the tail of the SM thermal distribution contributes to the production of DM; see, e.g. Refs.~\cite{Cosme:2023xpa, Koivunen:2024vhr, Arcadi:2024wwg, Arcadi:2024obp}. This approach corresponds to an {\it instantaneous} reheating. In the next section, instead, we study the genesis of DM {\it during} a non-vanishing reheating phase.

\subsection{During Reheating}
Alternatively, DM could also be produced during the reheating era, before the SM energy density domination. In the next subsections, the DM production through scatterings and decays will be analyzed.

\subsubsection{Through Scatterings} \label{sec:scattering}
\subsubsection*{Light DM}
In the case where DM is lighter than the temperature scales during reheating (that is $m < \Trh$ and $m < T_I$), the DM abundance can be computed by integrating Eq.~\eqref{eq:BE} between $a_I$ and $\arh$, so that
\begin{align}
    N_\text{rh} &= \int_{a_I}^{\arh} \frac{a^2}{H}\, \gamma_a\, da = \frac{\Trh^k}{\Hrh\, \Lambda^{k-4}} \int_{a_I}^{\arh} \frac{a^2}{(\arh/a)^{\frac32 (\omega+1)}} \left(\frac{\arh}{a}\right)^{\alpha\, k} da \nonumber\\
    &\simeq \frac{\Trh^k}{\Hrh\, \Lambda^{k-4}}\, \arh^3 \times
    \begin{dcases}
        \frac{1}{\alpha\, (k_c - k)} & \text{ for } \alpha\, k < \alpha\, k_c\,,\\
        \ln\frac{\arh}{a_I} & \text{ for } k = k_c\,,\\
        \frac{1}{\alpha\, (k - k_c)} \left(\frac{\arh}{a_I}\right)^{\alpha (k - k_c)} & \text{ for } \alpha\, k > \alpha\, k_c\,,
    \end{dcases}
\end{align}
where the critical value $k_c$ for $k$ is defined as
\begin{equation} \label{eq:kc}
    k_c \equiv \frac32\, \frac{\omega + 3}{\alpha}\,.
\end{equation}
Therefore, the DM yield at the end of reheating can be expressed as
\begin{equation} \label{eq:lightDM}
    Y_\text{rh} \simeq \frac{135}{2\pi^3\, \gss} \sqrt{\frac{10}{\gs}}\, \frac{M_P\, \Trh^{k-5}}{\Lambda^{k-4}} \times
    \begin{dcases}
        \frac{1}{\alpha\, (k_c - k)} & \text{ for } \alpha\, k < \alpha\, k_c\,,\\
        \frac{1}{\alpha}\, \ln\frac{T_I}{\Trh} & \text{ for } k = k_c\,,\\
        \frac{1}{\alpha\, (k - k_c)} \left(\frac{T_I}{\Trh}\right)^{k - k_c} & \text{ for } \alpha\, k > \alpha\, k_c\,.
    \end{dcases}
\end{equation}
Several comments are in order: $i)$ This result reproduces the behavior found in Ref.~\cite{Garcia:2017tuj} for a massive inflaton decaying into SM particles with a constant decay width (that is, $\omega = 0$ and $\alpha=3/8$), and in Ref.~\cite{Bernal:2019mhf}, for general equation-of-state parameters $\omega$ and $\alpha = \frac38 (1+\omega)$. $ii)$ The overall prefactor of Eq.~\eqref{eq:lightDM} has the same scaling shown in Eq.~\eqref{eq:DMrd}, however, new factors depending on $k$ appear. $iii)$ For $\alpha\, k < \alpha\, k_c$ an order one boost could occur, as the temperature dependence is identical. However, the ratio of DM production during and after reheating $\frac{\gss(T_0)}{\gss(\Trh)} \frac{k - 5}{(k_c - k)\, \alpha}$ is typically smaller than one, making DM production during reheating subdominant. $iv)$ In the case where $k = k_c$ the enhancement is logarithmic and larger than one, interdependently of the sign of $\alpha$. $v)$ A potentially large power-law boost can take place only if $\alpha\, k > \alpha\, k_c$. It occurs only for $\alpha > 0$, since $T_I > \Trh$ is required. $vi)$ We note that Eq.~\eqref{eq:lightDM} behaves well in the limit $\alpha = 0$.

\subsubsection*{Heavy DM with $\boldsymbol{\alpha > 0}$}
Alternatively, for $\alpha > 0$, one has $\Trh < T_I$, and then the hierarchy $\Trh < m < T_I$ becomes possible. In that case, Eq.~\eqref{eq:BE} can be integrated between $a_I$ and the scale factor $a_m$ at which $T=m$, so that
\begin{equation}
    N_\text{rh} \simeq \frac{\Trh^k}{\Hrh\, \Lambda^{k-4}}\, \arh^3 \times
    \begin{dcases}
        \frac{1}{\alpha\, (k_c - k)} \left(\frac{a_m}{\arh}\right)^{\alpha\, (k_c - k)} & \text{ for } k < k_c\,,\\
        \ln\frac{a_m}{a_I} & \text{ for } k = k_c\,,\\
        \frac{1}{\alpha\, (k - k_c)} \left(\frac{\arh}{a_I}\right)^{\alpha\, (k - k_c)} & \text{ for } k > k_c\,,
    \end{dcases}
\end{equation}
and therefore
\begin{equation} \label{eq:heavyDMpos}
    Y_\text{rh} \simeq \frac{135}{2\pi^3\, \gss} \sqrt{\frac{10}{\gs}}\, \frac{M_P\, \Trh^{k-5}}{\Lambda^{k-4}} \times
    \begin{dcases}
        \frac{1}{\alpha\, (k_c - k)} \left(\frac{\Trh}{m}\right)^{k_c - k} & \text{ for } k < k_c\,,\\
        \frac{1}{\alpha}\, \ln\frac{T_I}{m} & \text{ for } k = k_c\,,\\
        \frac{1}{\alpha\, (k - k_c)} \left(\frac{T_I}{\Trh}\right)^{k - k_c} & \text{ for } k > k_c\,.
    \end{dcases}
\end{equation}

\subsubsection*{Heavy DM with $\boldsymbol{\alpha < 0}$}
Finally, for $\alpha < 0$, one instead has $T_I < \Trh$. If $T_I < m < \Trh$, Eq.~\eqref{eq:BE} must be integrated between $a_m$ and $\arh$, so that
\begin{equation}
    N_\text{rh} \simeq \frac{\Trh^k}{\Hrh\, \Lambda^{k-4}}\, \arh^3 \times
    \begin{dcases}
        \frac{1}{\alpha\, (k_c - k)} & \text{ for } k > k_c\,,\\
        \ln\frac{\arh}{a_m} & \text{ for } k = k_c\,,\\
        \frac{1}{\alpha\, (k - k_c)} \left(\frac{\arh}{a_m}\right)^{\alpha\, (k - k_c)} & \text{ for } k < k_c\,,
    \end{dcases}
\end{equation}
and therefore
\begin{equation} \label{eq:heavyDMneg}
    Y_\text{rh} \simeq \frac{135}{2\pi^3\, \gss} \sqrt{\frac{10}{\gs}}\, \frac{M_P\, \Trh^{k-5}}{\Lambda^{k-4}} \times
    \begin{dcases}
        \frac{1}{\alpha\, (k_c - k)} & \text{ for } k > k_c\,,\\
        \frac{1}{\alpha}\, \ln\frac{m}{\Trh} & \text{ for } k = k_c\,,\\
        \frac{1}{\alpha\, (k - k_c)} \left(\frac{m}{\Trh}\right)^{k - k_c} & \text{ for } k < k_c\,.
    \end{dcases}
\end{equation}

\begin{figure}[t!]
    \def\sepf{0.496}
    \centering
    \includegraphics[width=\sepf\columnwidth]{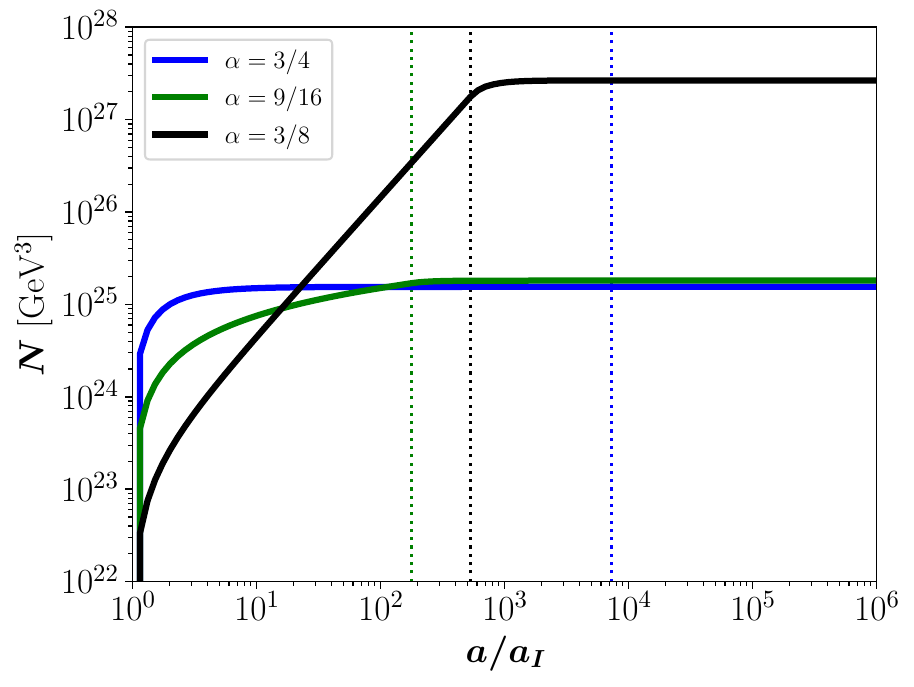}
    \includegraphics[width=\sepf\columnwidth]{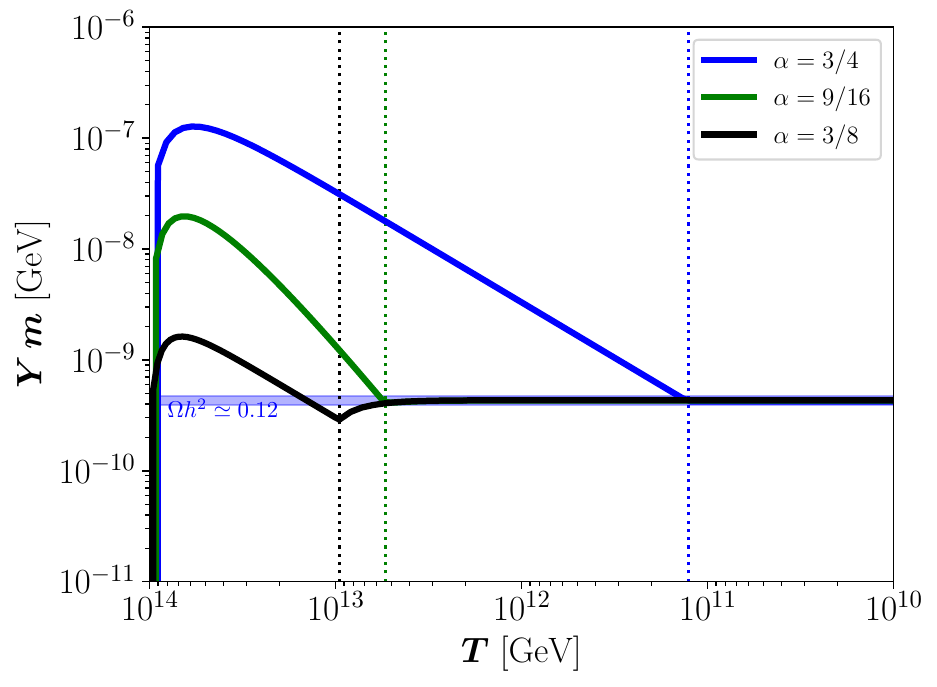}
    \caption{Evolution of the DM density as a function of the scale factor $a$ (left) or the temperature $T$ (right), for a DM production from the SM bath mediated by gravitons (cf. Section~\ref{sec:gravmediation}), for $m = 10^{12}$~GeV, $\omega = 0$, $T_I = 10^{14}$~GeV, $\alpha = 3/4$ and $\Trh \simeq 1.3 \times 10^{11}$~GeV (blue), $\alpha = 9/16$ and $\Trh \simeq 5.4 \times 10^{12}$~GeV (green), or $\alpha = 3/8$ and $\Trh \simeq 9.6 \times 10^{12}$~GeV (black). The vertical lines correspond to $a= \arh$ or $T=\Trh$. The horizontal band in the right panel corresponds to the observed DM abundance.}
    \label{fig:evolution}
\end{figure} 
The evolution of the DM density is shown in Fig.~\ref{fig:evolution}, as a function of the scale factor $a$ (left) or the temperature $T$ (right), for a DM production from the SM bath mediated by gravitons ($k=8$, cf. Section~\ref{sec:gravmediation}). The curves were produced by numerically solving Eq.~\eqref{eq:BE} in the background defined in Eqs.~\eqref{eq:Hubble} and~\eqref{eq:Tem}. We have assumed $m = 10^{12}$~GeV, $\omega = 0$, $T_I = 10^{14}$~GeV, $\alpha = 3/4$ and $\Trh \simeq 1.3 \times 10^{11}$~GeV ($k > k_c = 6$, blue), $\alpha = 9/16$ and $\Trh \simeq 5.4 \times 10^{12}$~GeV ($k = k_c = 8$, green), or $\alpha = 3/8$ and $\Trh \simeq 9.6 \times 10^{12}$~GeV ($k < k_c = 12$, black). The values for the reheating temperature were chosen to match the observed DM abundance (horizontal band on the right panel). In addition, the three choices for $\alpha$ allow one to explore the three possible hierarchies of $k_c$ with respect to $k$ previously explored. Interestingly, if $k < k_c$ (black lines) most of the DM is produced near $T \sim \Trh$, while if $k > k_c$ the bulk of the production occurs at the highest temperatures $T \sim T_I$ (blue lines). Furthermore, the effect of the DM dilution as a result of the injection of entropy during reheating can be clearly seen in the right panel. The vertical lines correspond to $a = \arh$ or $T = \Trh$.

\begin{figure}[t!]
    \def\sepf{0.496}
    \centering
    \includegraphics[width=\sepf\columnwidth]{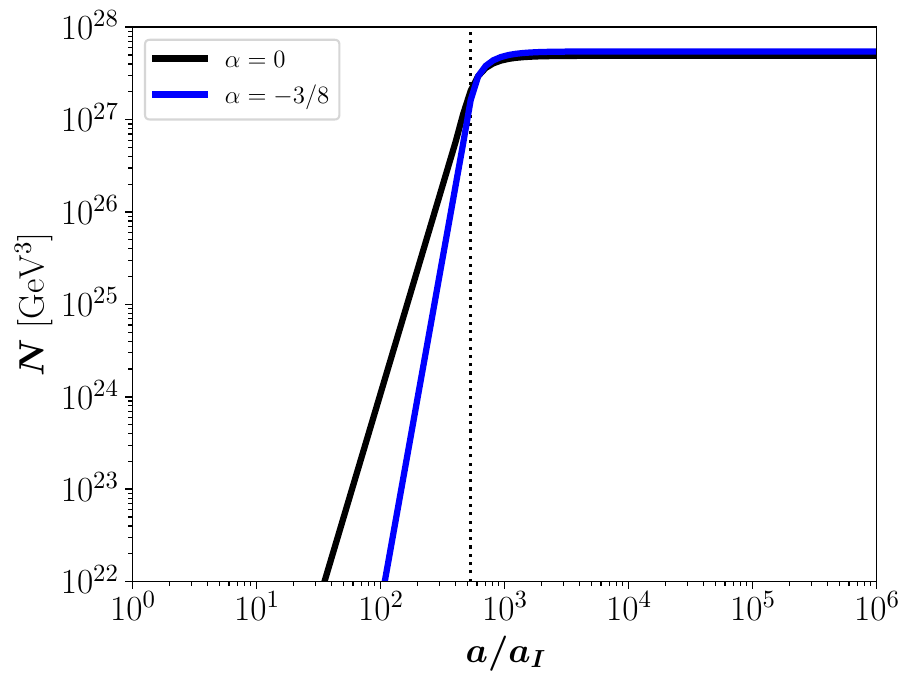}
    \includegraphics[width=\sepf\columnwidth]{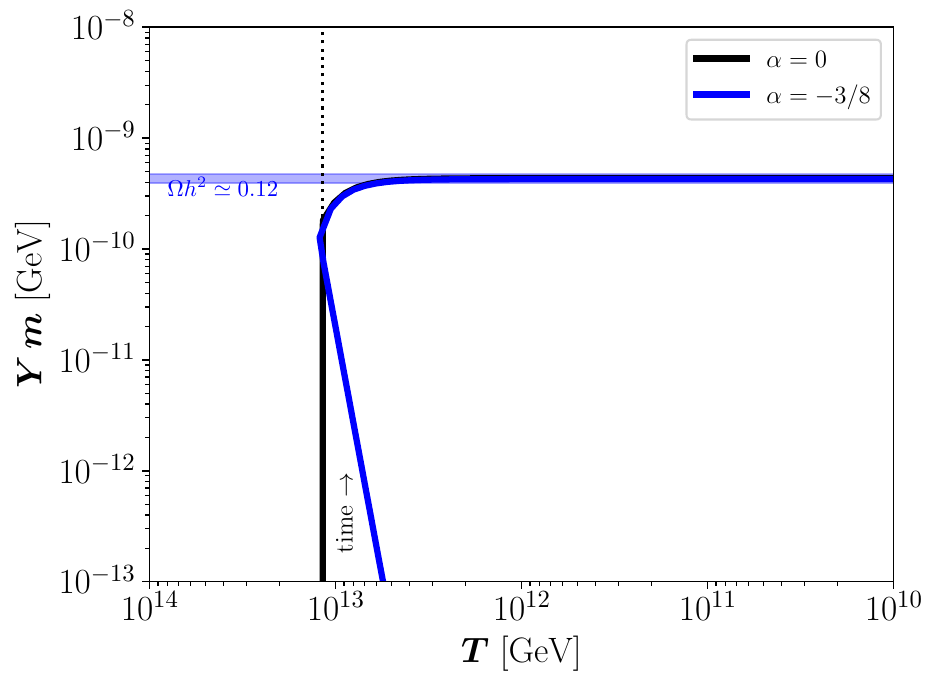}
    \caption{Same as Fig.~\ref{fig:evolution} but for $\Trh \simeq 1.2 \times 10^{13}$~GeV and $\alpha = 0$ (black) or $\alpha = -3/8$ (blue).}
    \label{fig:evolution2}
\end{figure}

For completeness, in Fig.~\ref{fig:evolution2} the cases $\alpha = 0$ (black) or $\alpha = -3/8$ (blue) are explored. For $\alpha = 0$, even if the SM temperature is constant during reheating, the universe is expanding and the DM abundance is slowly building up. The same applies to the case where $\alpha < 0$, but here the temperature increases during reheating. Both cases correspond to $k < k_c$, and therefore the DM is produced mainly at the end of reheating, when $T \simeq \Trh \simeq 1.2 \times 10^{13}$~GeV.

\subsubsection{Through Decays}
\subsubsection*{Light DM}
In the case of direct production of light DM (that is, $\mrh > x\, m$) from inflaton decays, taking into account the field dependence on the inflaton mass, cf. Eq.~\eqref{eq:mrh}, Eq.~\eqref{eq:BE} can be integrated from $a = a_I$ to $a = \arh$ which gives rise to
\begin{equation}
    Y_\text{rh} \simeq \frac34\, \frac{\gs}{\gss}\, \mathcal{C}\, \Trh \times
    \begin{dcases}
        4\, \frac{\alpha - 1}{4\alpha-3(1+\omega)} & \text{ for } \alpha < \frac34\, (1+\omega)\,,\\
        \frac{1-3\omega}{\alpha}\, \ln\frac{T_I}{\Trh} & \text{ for } \alpha = \frac34\, (1+\omega)\,,\\
        4\, \frac{\alpha-1}{3(1+\omega)-4\alpha} \left(\frac{T_I}{\Trh}\right)^\frac{4\alpha-3(1+\omega)}{\alpha} & \text{ for } \alpha > \frac34\, (1+\omega)\,.
    \end{dcases}
\end{equation}
We note that for $\alpha = \frac34\, (1+\omega)$, a logarithmic dependence on the ratio $T_I/\Trh$ appears. This dependence could become a power law if $\alpha > \frac34\, (1+\omega)$; however, it cannot be realized as it does not give rise to a viable reheating era; see Fig.~\ref{fig:alpha_omega}. Interestingly, the case of a constant inflaton mass can be recovered in the limit $\omega = 0$. 

\begin{figure}[t!]
    \def\sepf{0.496}
    \centering
    \includegraphics[width=\sepf\columnwidth]{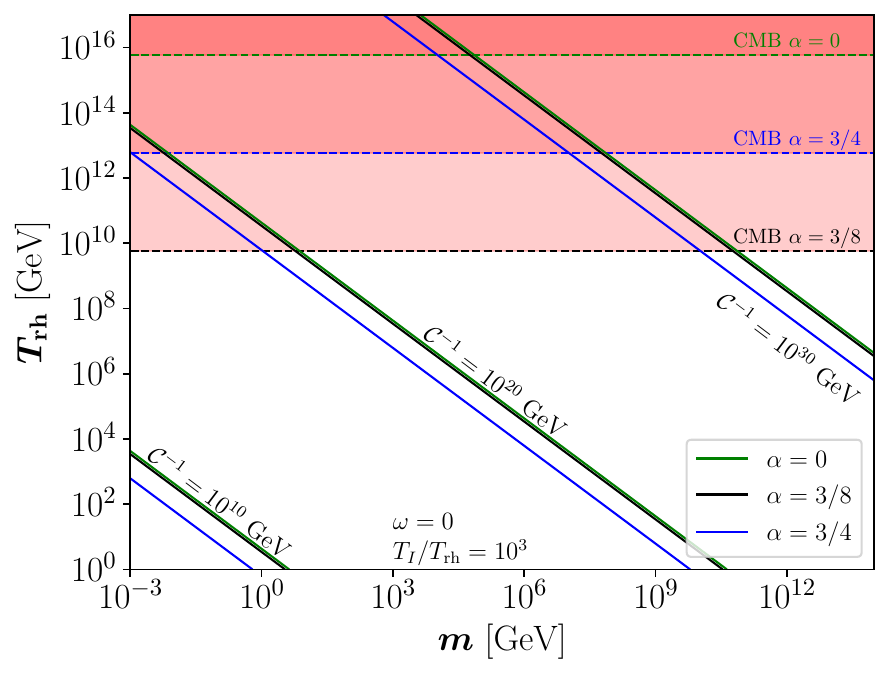}
    \caption{Parameter space required to fit the entire DM abundance through decays of the inflaton for $\omega = 0$, $\alpha = 0$ (green), $\alpha = 3/8$ (black) or $\alpha = 3/4$ (blue), and $\mathcal{C}^{-1} = 10^{10}$~GeV, $\mathcal{C}^{-1} = 10^{20}$~GeV or $\mathcal{C}^{-1} = 10^{30}$~GeV. The red areas correspond to the CMB constraint for the three values of $\alpha$.}
    \label{fig:decay}
\end{figure}

Figure~\ref{fig:decay} shows the parameter space required to fit the entire DM abundance through decays of the inflaton for $\omega = 0$, $\alpha = 0$ (green), $\alpha = 3/8$ (black) or $\alpha = 3/4$ (blue) and $\mathcal{C}^{-1} = 10^{10}$~GeV, $\mathcal{C}^{-1} = 10^{20}$~GeV or $\mathcal{C}^{-1} = 10^{30}$~GeV. Large values for $\mathcal{C}^{-1}$ can be achieved with large inflaton masses and very small branching fractions into DM particles. The curves corresponding to $\alpha = 0$ and $\alpha = 3/8$ almost overlap, as they both correspond to the regime $\alpha < \frac34 (1+\omega)$. However, the logarithmic enhancement for $\alpha = 3/4$ allows smaller values of $\Trh$ to be explored; in this case $T_I/\Trh = 10^3$ was chosen. The red areas correspond to the CMB constraint for the three values of $\alpha$; cf. Eq.~\eqref{eq:TI}.

\subsubsection*{Heavy DM}
Alternatively, for heavier DM with a mass in the range $\mrh < x\, m < m_\phi(a_I)$, Eq.~\eqref{eq:BE} must be integrated from $a = a_I$ to $a = a_\phi$ defined as
\begin{equation}
    a_\phi \equiv a(m_\phi = x\, m) = \arh \left(\frac{\mrh}{x\, m}\right)^\frac{1}{3\, \omega},
\end{equation}
for $\omega >0$. In this case,
\begin{equation}
    Y_\text{rh} \simeq \frac34\, \frac{\gs}{\gss}\, \mathcal{C}\, \Trh \times
    \begin{dcases}
        4\, \frac{\alpha - 1}{4\alpha-3(1+\omega)} \left(\frac{\mrh}{x\, m}\right)^{\frac23 \frac{3\omega-2\alpha}{\omega}} & \text{ for } \alpha < \frac34\, (1+\omega)\,,\\
        (1-3\omega)\, \ln\left[\left(\frac{T_I}{\Trh}\right)^\frac{1}{\alpha} \left(\frac{\mrh}{x\, m}\right)^\frac{1}{3\omega}\right] & \text{ for } \alpha = \frac34\, (1+\omega)\,,\\
        4\, \frac{\alpha-1}{3(1+\omega)-4\alpha} \left(\frac{T_I}{\Trh}\right)^\frac{4\alpha-3(1+\omega)}{\alpha} & \text{ for } \alpha > \frac34\, (1+\omega)\,.
    \end{dcases}
\end{equation}

\section{Dark Matter Gravitational Production} \label{sec:particles}
In this section, we explore two examples of DM UV freeze-in production through scatterings: first annihilations of SM particles during and after reheating, and then annihilations of inflaton during reheating. We note that this case where the DM only interacts gravitationally with the SM is, by construction, a nightmare scenario to probe experimentally.\footnote{The proposed Windchime experiment~\cite{Windchime:2022whs} is sensitive to purely gravitational DM, as long as its mass is close to or larger than the Planck scale.} However, it is interesting to emphasize that gravity is the only irreducible force between the visible and the dark sectors.

\subsection{SM scatterings} \label{sec:gravmediation}
DM particles are unavoidably produced in annihilations of SM particles from the thermal bath, mediated by the $s$-channel exchange of gravitons. The corresponding interaction rate is
\begin{equation} \label{eq:gravity}
    \gamma_a \simeq \mathcal{C}_g\, \frac{T^8}{M_P^4}\,,
\end{equation}
where $\mathcal{C}_g \simeq 1.9 \times 10^{-4}$ for scalar DM, $\mathcal{C}_g \simeq 1.1 \times 10^{-3}$ for fermionic DM, or $\mathcal{C}_g \simeq 2.3 \times 10^{-3}$ for vector DM, summing over the contributions from all SM degrees of freedom in the thermal plasma~\cite{Garny:2015sjg, Tang:2017hvq, Garny:2017kha, Bernal:2018qlk, Barman:2021ugy}. In the notation of Eq.~\eqref{eq:gammaa}, this corresponds to $k=8$ and $\Lambda = M_P/\mathcal{C}_g^{1/4}$, and therefore the results of Sections~\ref{sec:after} and~\ref{sec:scattering} can be used directly. Examples of the evolution of the DM comoving number density were shown in Figs.~\ref{fig:evolution} and~\ref{fig:evolution2}.

\begin{figure}[t!]
    \def\sepf{0.496}
    \centering
    \includegraphics[width=\sepf\columnwidth]{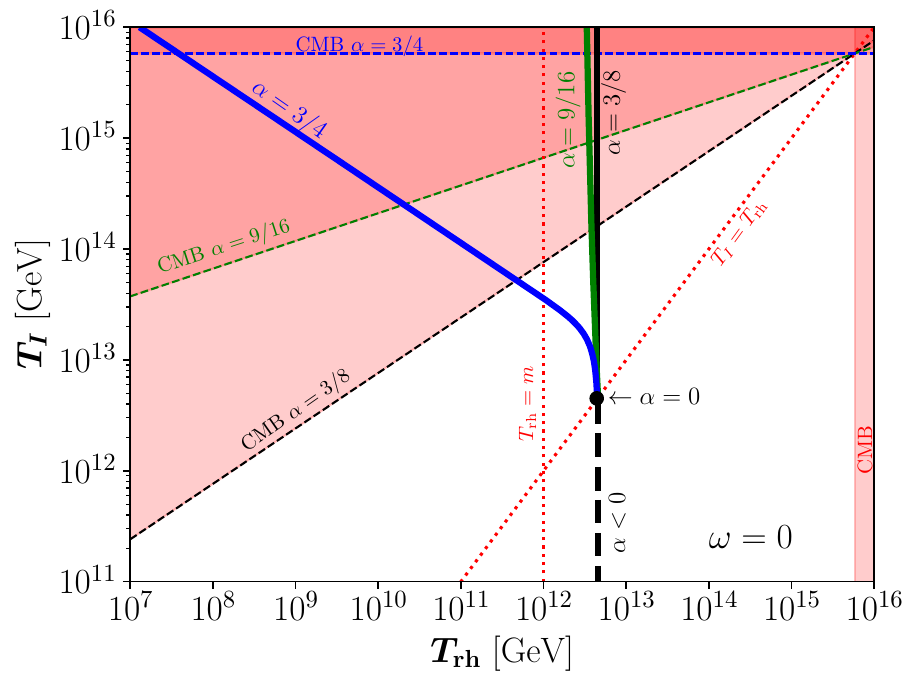}
    \includegraphics[width=\sepf\columnwidth]{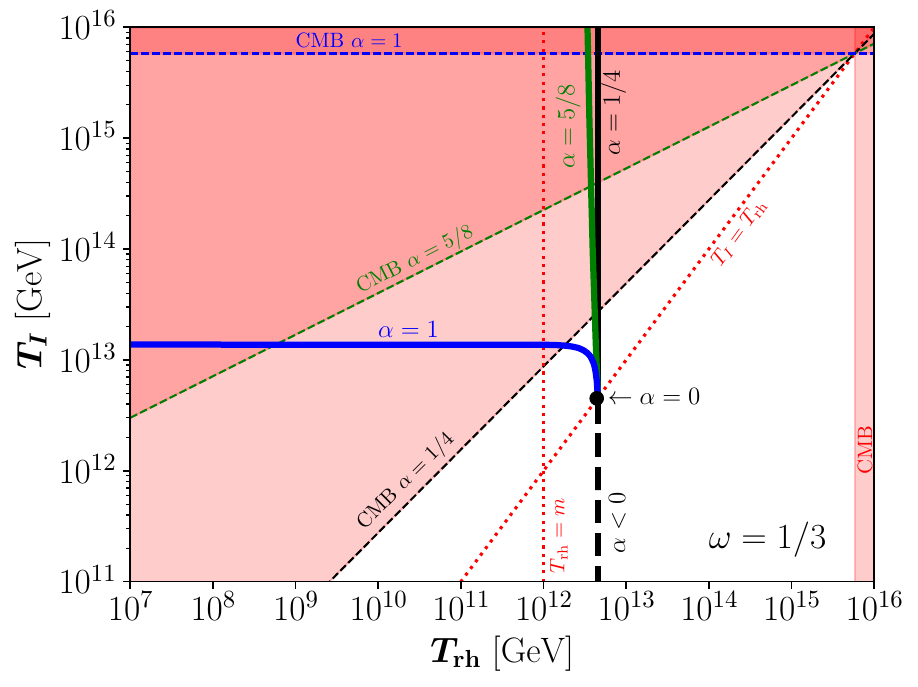}
    \caption{Parameter space required to fit the entire DM abundance through gravitational annihilation of SM particles for $m = 10^{12}$~GeV, $\omega = 0$ (left) or $\omega =1/3$ (right), and different values of $\alpha$, for fermionic DM. The dotted red lines correspond to $T_I = \Trh$ and $\Trh = m$. The red areas correspond to the CMB constraint for the three values of $\alpha$.}
    \label{fig:gravity}
\end{figure} 
The left panel of Fig.~\ref{fig:gravity} shows the parameter space required to fit the entire DM abundance through gravitational annihilation of SM particles, that is, Eq.~\eqref{eq:gravity}, taking into account both the contributions after and during reheating, for fermionic DM. In the figure, $\omega = 0$ and $m = 10^{12}$~GeV have been used, taking $\alpha = 3/4$ (solid blue), $\alpha = 9/16$ (solid green), $\alpha = 3/8$ (solid black), $\alpha = 0$ (black dot) and $\alpha < 0$ (dashed black). The first three choices of $\alpha$ give rise to $k_c = 12$, $k_c = 8$, and $k_c = 6$, respectively, and hence to the three cases described in Eqs.~\eqref{eq:lightDM} and~\eqref{eq:heavyDMpos}. For reference, the lines corresponding to these three choices of $k_c$ are shown in Fig.~\ref{fig:alpha_omega}. For $\alpha = 3/8$, one has $k < k_c$: DM is produced mainly at the end of reheating, and therefore its abundance is insensitive to the value of $T_I$, which translates as a vertical line. However, a small logarithmic dependence on $T_I$ appears when $\alpha = 9/16$ as $k = k_c$: the line slightly bends towards smaller values of $\Trh$. Finally, for $\alpha = 3/4$, $k$ goes above its critical value $k > k_c$, and therefore the abundance of DM receives a boost proportional to $(T_I/\Trh)^2$ that allows the exploration of much smaller values of $\Trh$. As expected, the three lines merge when $T_I$ approaches $\Trh$, as the reheating tends to be instantaneous. This merging point also corresponds to the case in which $\alpha = 0$, where $T_I = \Trh$. For $\alpha < 0$, the temperature increases during reheating, so that $T_I < \Trh$, the dependence on $T_I$ is very suppressed and, therefore, the parameter space required to match the whole observed DM abundance becomes a vertical line. In addition, the red areas correspond to the CMB constraint on the inflationary scale for different values of $\alpha$; cf. Eq.~\eqref{eq:TI}. It is interesting to note that larger values of $\alpha$ allow us to reach higher values of $T_I$. In the present case with $\omega = 0$, one has to have $\alpha \leq 3/4$ to have viable reheating, cf. Fig.~\ref{fig:alpha_omega}.

The right panel of Fig.~\ref{fig:gravity} compares to the left panel, but for $\omega = 1/3$, taking $\alpha = 1$ (solid blue), $\alpha = 5/8$ (solid green), $\alpha = 1/4$ (solid black), $\alpha = 0$ (black dot) and $\alpha < 0$ (dashed black). These first three choices of $\alpha$ give rise to $k_c = 5$, $k_c = 8$, and $k_c = 20$, respectively. It is interesting to note that for $\alpha = 1$ one has $k_c = 5$, and therefore Eq.~\eqref{eq:heavyDMpos} becomes independent of $\Trh$, which is reflected in the plot by a horizontal line.

\subsection{Inflaton scatterings} \label{sec:inflaton}
Alternatively, during reheating DM can also be produced by scatterings of inflatons mediated by the exchange of gravitons; the corresponding interaction-rate density for fermionic DM is given by~\cite{Mambrini:2021zpp, Bernal:2021kaj, Barman:2021ugy, Barman:2022tzk, Bernal:2024ykj}
\begin{equation}
    \gamma \simeq \frac{\rp^2}{m_\phi^2}\, \frac{m^2}{64\pi\, M_P^4} \simeq \frac{9}{64\pi}\, \frac{m^2\, \Hrh^4}{\mrh^2} \left(\frac{T}{\Trh}\right)^\frac{6}{\alpha},
\end{equation}
where the dependence in $\omega$ cancels out in the ratio $\rp/m_\phi$. Again, in the notation of Eq.~\eqref{eq:gammaa}, this corresponds to
\begin{equation}
    \Lambda = \Trh \left(\frac{64\pi}{9}\, \frac{\mrh^2\, \Trh^4}{m^2\, \Hrh^4}\right)^\frac{\alpha}{2(3-2\alpha)},
\end{equation}
and $k = 6/\alpha$. As in this case where $k$ is no longer constant and depends on $\alpha$, the equivalent of the critical exponent $k_c$ becomes a critical equation-of-state parameter $\omega_c = 1$. Even if the general expressions in Section~\ref{sec:scattering} could be easily translated to this case, for the sake of clarity, we prefer to explicitly report how Eq.~\eqref{eq:lightDM} reflects to this case:
\begin{equation} \label{eq:inflaton}
    Y_\text{rh} \simeq \frac{1}{128}\, \sqrt{\frac{\gs}{10}}\, \frac{\gs}{\gss} \frac{\Trh^3\, m^2}{M_P^3\, \mrh^2} \times
    \begin{dcases}
        \frac{1}{1-\omega} \left(\frac{T_I}{\Trh}\right)^{\frac32 \frac{1-\omega}{\alpha}} & \text{ for } \omega < 1\,,\\
        \frac{3}{2 \alpha}\, \ln\frac{T_I}{\Trh} & \text{ for } \omega = 1\,,\\
        \frac{1}{\omega-1} & \text{ for } \omega > 1\,.
    \end{dcases}
\end{equation}

\begin{figure}[t!]
    \def\sepf{0.496}
    \centering
    \includegraphics[width=\sepf\columnwidth]{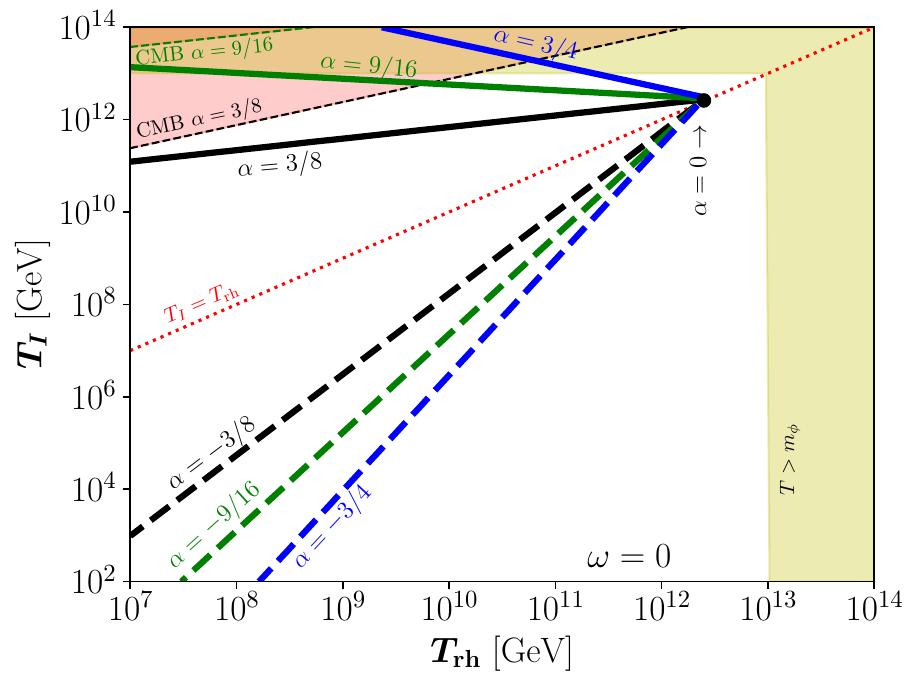}
    \includegraphics[width=\sepf\columnwidth]{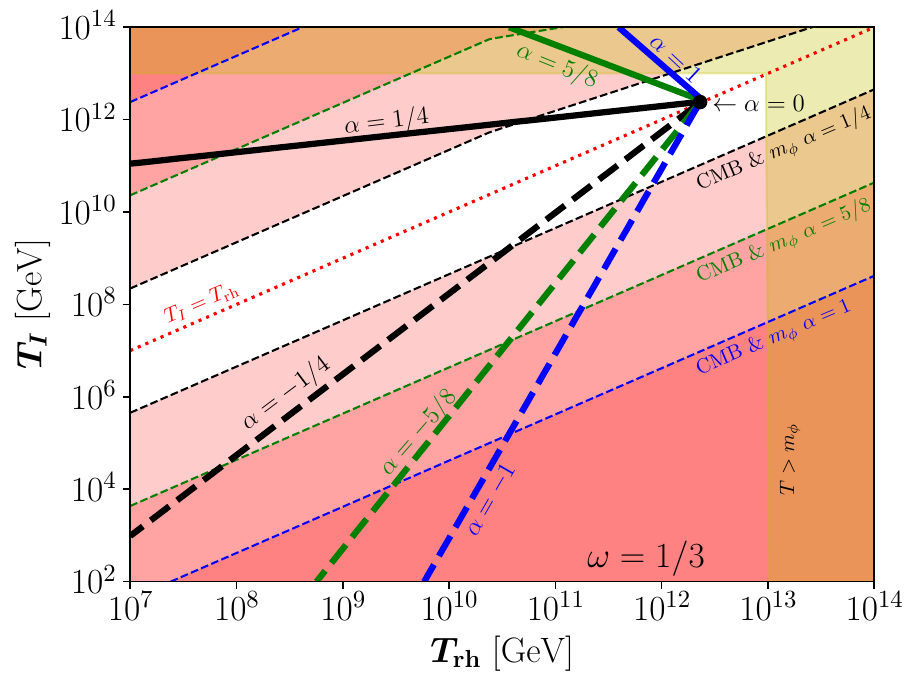}
    \caption{Parameter space required to fit the entire DM abundance through gravitational annihilation of inflatons for $m = 10^{12}$~GeV, $\mrh = 10^{13}$~GeV, $\omega = 0$ (left) or $\omega =1/3$ (right), and different values of $\alpha$, for fermionic DM. The red dotted lines correspond to $T_I = \Trh$. The yellow areas are excluded because the temperature is higher than the inflaton mass. The red areas correspond to the the combination of the constraints from CMB and the inflaton mass, for the three values of $\alpha$.}
    \label{fig:inflaton}
\end{figure}

The left panel of Fig.~\ref{fig:inflaton} shows the parameter space required to fit the entire DM abundance through gravitational annihilation of inflatons for $m = 10^{12}$~GeV, $\mrh = 10^{13}$~GeV, and $\omega = 0$, taking $|\alpha| = 3/4$ (blue), $|\alpha| = 9/16$ (green), $|\alpha| = 3/8$ (black) and $\alpha = 0$ (black dot). It only takes into account the contribution during reheating, where inflatons dominate the energy density of the Universe. The positive and negative values for $\alpha$ are depicted with solid and dashed lines, respectively. The red dotted line corresponds to $T_I = \Trh$. The yellow areas are excluded because the temperature is higher than the inflaton mass, while the red areas correspond to the CMB constraint on the inflationary scale for the three values of $\alpha$. This choice for the equation-of-state parameter is described by the first case of Eq.~\eqref{eq:inflaton} (e.g. $\omega  < 1$) and therefore a strong dependence on the ratio $T_I/\Trh$ is expected, as can be observed in the figure. The different lines converge when $T_I = \Trh$, corresponding to an instantaneous reheating era, and to the case $\alpha = 0$ where the temperature is constant during reheating. For $\alpha >0$, very large values of $T_I$ are excluded due to the constraint on the inflationary scale. However, for $\alpha < 0$, $T_I < \Trh$ and much smaller values of $T_I$ can give rise to the abundance of DM observed. Interestingly, and in contrast to the previous case of scattering of SM particles, here both positive and negative values for $\alpha$  give rise to a power-law enhancement.

The right panel of Fig.~\ref{fig:inflaton} compares to the left panel, but for $\omega = 1/3$, taking $|\alpha| = 1$ (blue), $|\alpha| = 5/8$ (green) and $|\alpha| = 1/4$ (black). For this case, the parameter space that reproduces the observed DM abundance is generally similar; however, the constraint on the inflaton mass becomes more involved, since $m_\phi$ is no longer constant. The red bands correspond to the combination of the constraints from the CMB and the inflaton mass. For $T_I > \Trh$ a stronger upper bound appears, coming from Eq.~\eqref{eq:TImax}, while for $T_I < \Trh$ a lower bound from Eq.~\eqref{eq:TImin} exists for the three values of $\alpha$.

\section{Conclusions} \label{sec:conclusion}
In this work, we have studied the phenomenology of dark matter (DM) production in the early Universe through the ultraviolet (UV) freeze-in (FIMP) paradigm. In this scenario, the bulk of the DM is generated at the highest temperatures reached by the standard model (SM) thermal plasma, typically during the cosmic reheating era. Taking into account that the dynamics of reheating is largely unknown, we used a general parametrization for the Hubble expansion rate and SM temperature as a function of the cosmic scale factor, to bracket the uncertainty of the reheating epoch. This approach allows to recover standard reheating scenarios such as the cases where $i)$ the inflaton oscillates at the bottom of a monomial potential while decaying or annihilating into SM states, $ii)$ the inflaton decays with a general field-dependent decay width, $iii)$ the SM temperature during reheating is constant, $iv)$ or even scenarios where the energy density inflaton of the inflaton gets diluted faster than radiation so that its decay is not required. However, we emphasize that this approach also allows the exploration of a wide variety of alternative reheating histories, as long as the Hubble expansion and the SM temperature behave as simple power laws.

Within this approach to reheating, we analytically computed the production of DM in the early Universe from decays and annihilations, using a general temperature dependence for the interaction rates. Depending on the details of the reheating era and the DM production rate, the DM production could peak at the beginning or the end of reheating, and the final DM relic abundance could have a logarithmic or a power-law dependence on the ratio of temperatures at the beginning and the end of reheating. We explored the parameter space required to match the observed DM abundance, bracketing systematic uncertainties related to the reheating dynamics. Previous results reported in the literature were recovered, while viable alternative reheating evolutions were also investigated. This generic setup was used in the context of two particularly well-motivated examples: The gravitational DM production from the scattering of SM particles and scatterings of inflatons.

Before closing, we note that the present scenario is challenging to test experimentally for several reasons. $i)$ Due of its very suppressed interactions with the visible sector, the UV freeze-in paradigm evades detection with standard techniques. $ii)$ The observables sensitive to the cosmic reheating period are scarce. However, novel detectors such as the proposed Windchime experiment~\cite{Windchime:2022whs} are sensitive to heavy DM, even if it only interacts gravitationally with the SM. Additionally, gravitational waves provide a promising method for probing the early Universe and, in particular, the reheating era~\cite{Ghiglieri:2015nfa, Ghiglieri:2020mhm, Ringwald:2020ist, Muia:2023wru, Villa:2024jkw, Frey:2024jqy, Villa:2024jbf, Bernal:2024jim, Xu:2024cey}. As we look forward, we are excited to see progress in our theoretical grasp of cosmic reheating, which can enhance our comprehension of the early Universe and offer fresh perspectives on the nature of DM.

\acknowledgments
The authors thank Sasha Pukhov for helpful conversations and Yong Xu for comments on the draft. NB received funding from the Grant PID2023-151418NB-I00 funded by MCIU/AEI/10.13039/501100011033/ FEDER, UE.

\bibliographystyle{JHEP}
\bibliography{biblio}
\end{document}